\documentclass[sigconf]{acmart} 

\AtBeginDocument{%
  \providecommand\BibTeX{{%
    \normalfont B\kern-0.5em{\scshape i\kern-0.25em b}\kern-0.8em\TeX}}}

\setcopyright{acmlicensed}
\copyrightyear{2024}
\acmYear{2024}

\acmConference[RecSys'24]{Temporal Reasoning in Recommender Systems Workshop}{Oct 14--18, 2024}{Bari, Italy}
\acmPrice{15.00}
\acmISBN{978-1-4503-XXXX-X/18/06}

\usepackage[ruled,vlined]{algorithm2e}
\usepackage{soul}

\begin{document}

\title{CoActionGraphRec: Sequential Multi-Interest Recommendations Using Co-Action Graphs}

\author{{Yi Sun}}
\affiliation{%
  \institution{{eBay Inc.}}
  \country{{China}}
}
\email{{ysun24@ebay.com}}

\author{{Yuri M. Brovman}}
\affiliation{%
  \institution{{eBay Inc.}}
  \country{{USA}}
}
\email{{ybrovman@ebay.com}}

\renewcommand{\shortauthors}{{Y. Sun, et al.}}

\begin{abstract}
    There are unique challenges to developing item recommender systems for e-commerce platforms like eBay due to sparse data and diverse user interests. While rich user-item interactions are important, eBay's data sparsity exceeds other e-commerce sites by an order of magnitude. To address this challenge, we propose CoActionGraphRec (CAGR), a text based two-tower deep learning model (Item Tower and User Tower) utilizing co-action graph layers. In order to enhance user and item representations, a graph-based solution tailored to eBay's environment is utilized. For the Item Tower, we represent each item using its co-action items to capture collaborative signals in a co-action graph that is fully leveraged by the graph neural network component. For the User Tower, we build a fully connected graph of each user's behavior sequence, with edges encoding pairwise relationships. Furthermore, an explicit interaction module learns representations capturing behavior interactions. Extensive offline and online A/B test experiments demonstrate the effectiveness of our proposed approach and results show improved performance over state-of-the-art methods on key metrics.
\end{abstract}



\maketitle

\section{Introduction}\label{sec:introduction}

Item recommender systems have become an indispensable part of e-commerce platforms, playing a crucial role in connecting users with relevant items that they may be interested in. The eBay e-commerce platform has 2 billion live listings (items) and 132M active buyers (users). From a recommender system perspective, this is a very sparse dataset. Furthermore, in addition to the cold start problem, many items are volatile: many new items are listed daily and some items expire without accumulating substantial user interaction. This paper presents an approach for generating personalized item recommendations in this challenging data environment.

In general, similar to information retrieval (IR) systems, recommender systems operate in two stages: candidate generation and ranking~\cite{Covington2016}. The candidate generation stage efficiently retrieves a set of relevant items from the large item corpus, while the ranking stage sorts the items optimizing a business relevant metric such as user engagement or conversion. In this paper, we will focus on the candidate generation stage. Personalizing the candidate generation stage with user information using sequence-based recommendation has attracted significant research interest~\cite{Sun2019,Chang2021,Yu19,Li17,Chen22}. Additionally, multi-interest learning methods that capture users' diverse interests from their behavior sequences have shown promising results for long term user sequential modeling~\cite{Cen2020,Li2019,Tian22,Tan21}, and it has become one of the common industrial solutions for the candidate generation stage.

While existing sequential recommendation techniques form a solid foundation, there are still some limitations which we found that could be improved during our development of an e-commerce industrial solution.

\begin{figure*}[t]
    \small
    \centering
    \begin{minipage}{0.90\textwidth}
        \includegraphics[width=\columnwidth]{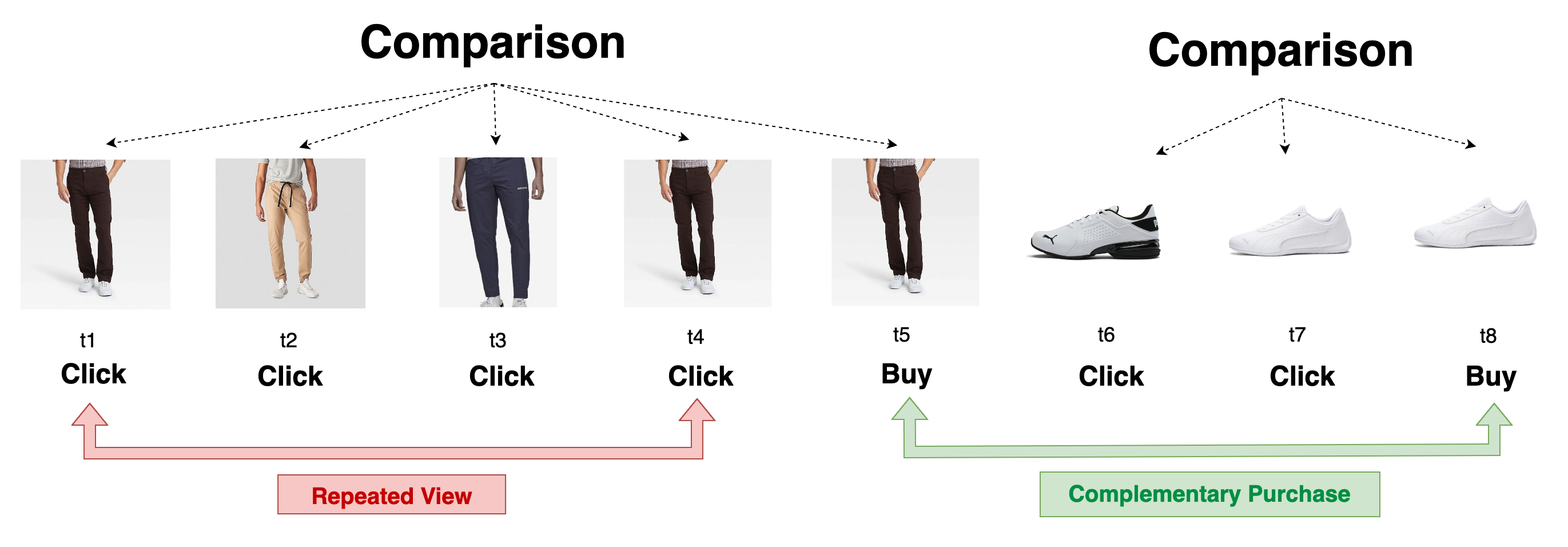}
    \end{minipage}
    \hfill
    \caption{User behavior sequence example. The user initially wanted to buy a pair of pants. After comparing multiple items, he viewed one of the pants at two time points: t1 and t4~(i.e. "Repeated View"), and then decided to purchase the pair of pants. Then he thought of pairing the pants with a pair of shoes, so he continued browsing and selected a suitable pair of shoes to place an order (i.e. "Complementary Purchase").} 
    \label{cmp}
\end{figure*}

\textbf{Leverage co-action items.} Sparsity is a common phenomenon in behavioral data of item recommender systems, particularly evident in some industrial-level scenarios. Due to extreme sparsity, we lack sufficient collaborative filtering signals to accurately calculate the relevance between a given user and an item. In order to make better use of the already sparse interaction behavior, we supplement the target item with its co-action items as additional features on the Item Tower of our CAGR deep learning model. These co-action items refer to item pairs in which users have interacted with both items. We employed a graph neural network (GNN) structure to fully leverage the co-action graph for enhanced item representations. During training, we employ a dual-objective loss to tune the importance of collaborative aspects.

\textbf{Explicit sequential interaction modeling.} While existing multi-interest learning methods have shown promise in extracting user interests from behavior sequences, they lack explicit modeling of interactions between user actions. It is quite common that users make purchase decisions after extensive product comparisons considering factors like item aspects, seller quality, price, etc., and it often results in repeated views of same items as well. Thus, modeling the behavior interaction by explicitly comparing item attributes between two user action items in a pairwise manner can better reflect comparison processes in shopping activities. For example, Figure~\ref{cmp} depicts a user shopping process, where the user demonstrates interest in pants and buy a dark color pants in the end, what we hope to do by explicit interaction modeling is to capture the signal that the user viewed items with the gray color in all the pants steps, except $t2$. Additionally, users may exhibit diverse actions like view, watch, add to cart, and purchase for a single item. While previous multi-behavior recommendation works like~\cite{Jin2020} and~\cite{Xia2021} have leveraged multi-behavior data, we can further exploit its potential by explicit pairwise modeling as well. For example, a user's clicks on different items may indicate the similarity relationship between them, while his purchases may indicate the complementary relationship between those bought items. So if we only use the click action which is the easiest action type to be collected, it will be more difficult to capture the patterns like complementary purchase, since it happens only when a user has at least two item purchases occur in a behavior sequence. In Figure~\ref{cmp} the user bought a shoes to match the pants that he just bought, which is the example of complementary purchase. 

To achieve the explicit pairwise modeling, we build a fully connected graph based on each user's behavior sequence, with direct edges between any two actions encoding their relationship. Based on this sequence graph, we applied another GNN module to learn representations for each node, capturing interactions between behaviors with pairwise edge information. In the above example, given each action pair we can directly capture the signal like whether the color are same, or the actions are the same type(e.g. both purchase), by storing the pairwise information on the edge. This is not easy to achieve for non-graph sequential modeling approaches like standard Transformer architecture, as it do not have the edge to store information with specific aspects of each action item. In standard Transformer, the items can only be transformed into a single, unified embedding to complete the sequential interaction by self-attention.

To summarize, in this work we have the following key contributions for improving multi-interest sequence behavior recommendation models:

\begin{itemize}
    \item For the Item Tower side, user co-action data is added in the form of a GNN component, which improves on the limitations of previous sequence recommendation models that only added user behavioral data to the User Tower side.
    \item For the User Tower side, we propose a general module for learning the pairwise interaction relationships (similarity, complementary, etc.) of user behavior. 
    \item A novel explicit interaction algorithm that expands on the standard GNN aggregation style and leverages the edge information as part of the aggregation source.
    \item The model has been deployed to full production traffic in a high scale e-commerce online marketplace, showing significant performance improvement in operational metrics.
\end{itemize}

\begin{figure*}[t]
    \small
    \centering
    \begin{minipage}{0.90\textwidth}
        \includegraphics[width=\columnwidth]{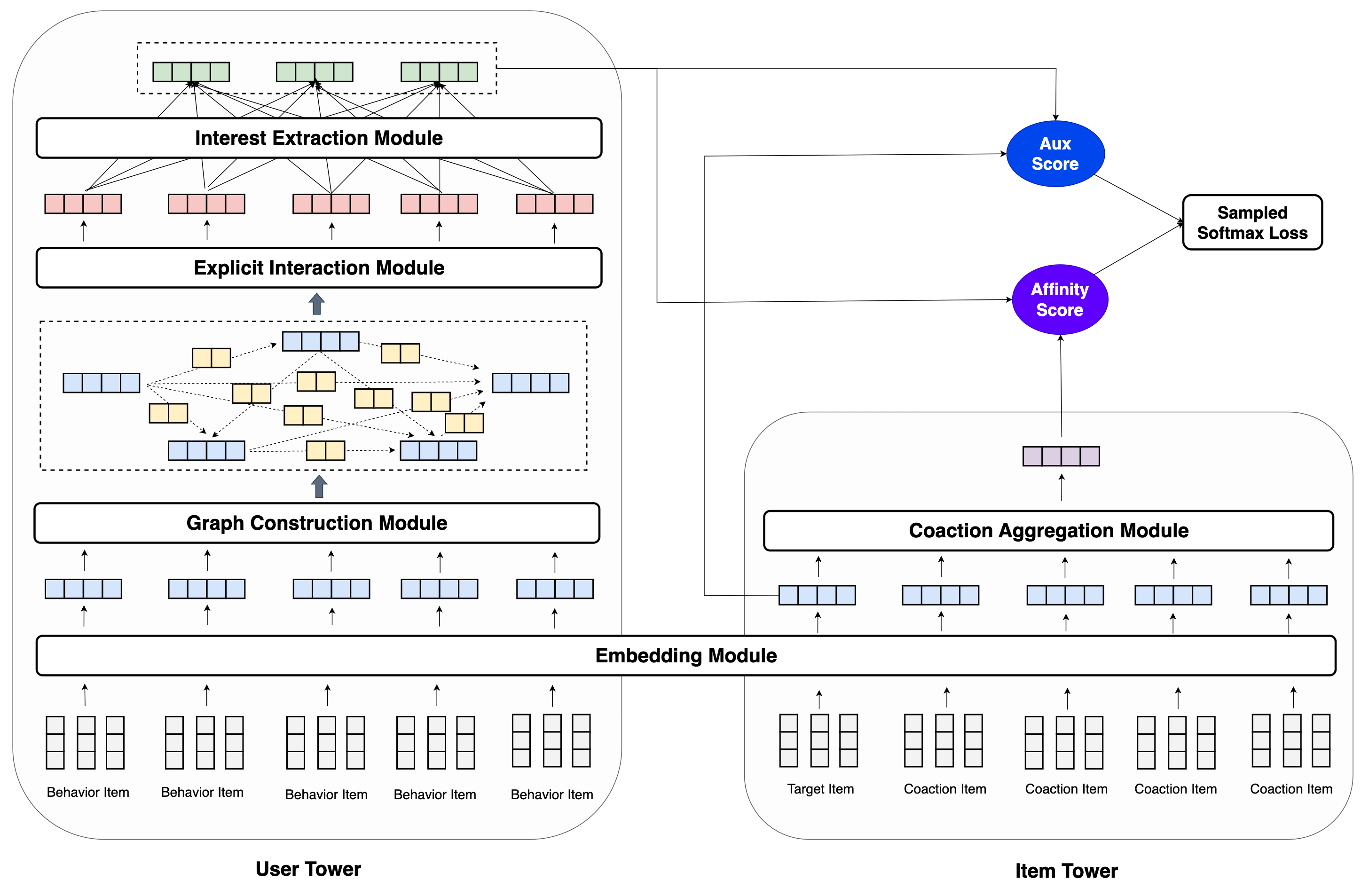}
    \end{minipage}
    \hfill
    \caption{Model Architecture Overview.}
    \label{model}
\end{figure*}

\section{Related Work}\label{sec:related-work}

\noindent \textbf{Multi-Interest Learning}. Typical sequential recommendation systems often model users' preferences as a single low-dimensional vector, which is not sufficient to accurately capture the diverse set of interests of a user. Recognizing this limitation, MIND~\cite{Li2019} introduced a dynamic routing mechanism that aggregates users' historical behaviors into multiple interest capsules. Building upon this work, ComiRec~\cite{Cen2020} further explored multi-head attention-based multi-interest routing to capture users' diverse interests and introduced diversity controllable methods. PIMIRec~\cite{ijcai2021p197} and UMI~\cite{Chai2022} further incorporated additional factors such as time information, interactivity, and user profiles. Additionally, Re4~\cite{Zhang2022} and REMI~\cite{Xie2023} optimized the training process by considering new loss terms and negative sampling strategies.

\noindent \textbf{Graph Neural Network for Recommendation}. Graph Neural Networks (GNNs) have emerged as a powerful tool for recommendation systems, enabling more accurate and personalized recommendations. PinSage~\cite{Ying2018} introduced the application of Graph Convolutional Networks (GCN) to pin-board graphs, marking the first work to incorporate GCN into industrial recommender systems. LightGCN~\cite{He2020} employed matrix factorization to derive user and item embedding matrices for recommendations, going beyond collaborative filtering. SR-GNN~\cite{Wu2019} stands out as one of the earliest works to utilize GNN for sequential data analysis. GCE-GNN~\cite{Wang2020} further enhanced session-based modeling by incorporating both global and session-level graph information. SURGE~\cite{Chang2021} employed GNN to address the challenges posed by noisy signals and preference drifting in long behavior sequences.

\noindent \textbf{Multi-Behavior Recommendation}. The concept of multi-behavior recommendation recognizes the importance of considering different types of user behaviors to gain a comprehensive understanding of user preferences. Multi-behavior recommendation often leverage Graph Neural Network to extract the valued information. MBGCN~\cite{Jin2020} proposes a multi-behavior recommendation model based on graph convolutional networks, capturing complex dependencies between different types of user behaviors. MBGMN~\cite{Xia2021} additionally utilizes a meta-learning framework to capture the relationships between different types of user behaviors. Besides Graph Neural Network, the transformer based architecture is suitable for leverage multi-behavior signals as well. MBSTR~\cite{Yuan2022} presents a multi-behavior recommendation model based on the sequential transformer architecture.

\begin{figure*}[t]
    \small
    \centering
    \begin{minipage}{0.95\textwidth}
        \includegraphics[width=\columnwidth]{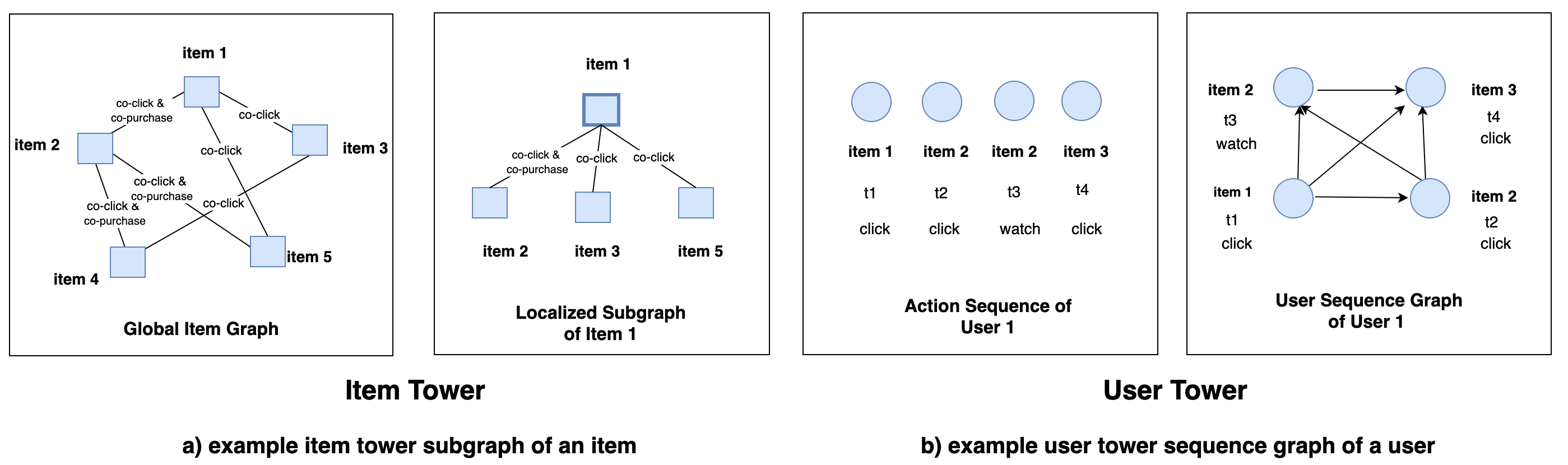}
    \end{minipage}
    \hfill
    \caption{Example of Item Tower and User Tower graph construction. a) The global item graph has 5 items with edges representing the co-actions. We extract the subgraph of item 1 from the global item graph. Here $\{$node = eBay item, edge = co-action$\}$. b) User sequence with 4 steps. Each step corresponds to a node in the user graph. Each node represents a user action on an eBay item at time $t_i$, with $t_1 < t_2 < t_3 < t_4$. The sequence is converted to a directed graph of 4 nodes. The directed relationship is determined by temporal order, since a user action can only be influenced by past actions, not future actions. Here $\{$node = user action on an eBay item, edge = temporal direction (edge weight learned during model training)$\}$. }
    \label{graph}
\end{figure*}

\section{Methodology}

\subsection{Problem Formulation}

Assume we have a set of users $u \in \mathcal{U}$ and each user $u$ has a sequence of actions $s_{u_1}, \ldots, s_{u_T}$, where $s_{u_i} = (x_{u_i}, b_{u_i}, t_{u_i})$, $x_{u_i}$ represents the item of $i$th action, $b_{u_i}$ is the behavior type of $i$th action, and $t_{u_i}$ is the behavior timestamp of $i$th action. In addition, given an item $x_q$, we define two types its co-action items: $x_{qc} \in$ $ \mathcal{C}_q $ when there is at least one user have click on both $x_q$ and $x_{qc}$, which we will refer to as co-click items; $x_{qp} \in$ $ \mathcal{P}_q $ when there is at least one user have purchase on both $x_q$ and $x_{qp}$, which we will refer to as co-purchase items. Our goal is to predict the next action of the user from a set of items $i \in \mathcal{I}$.

\subsection{Model Architecture}\label{subsec:architecture}

\begin{figure*}[t]
    \small
    \centering
    \begin{minipage}{0.95\textwidth}
        \includegraphics[width=\columnwidth]{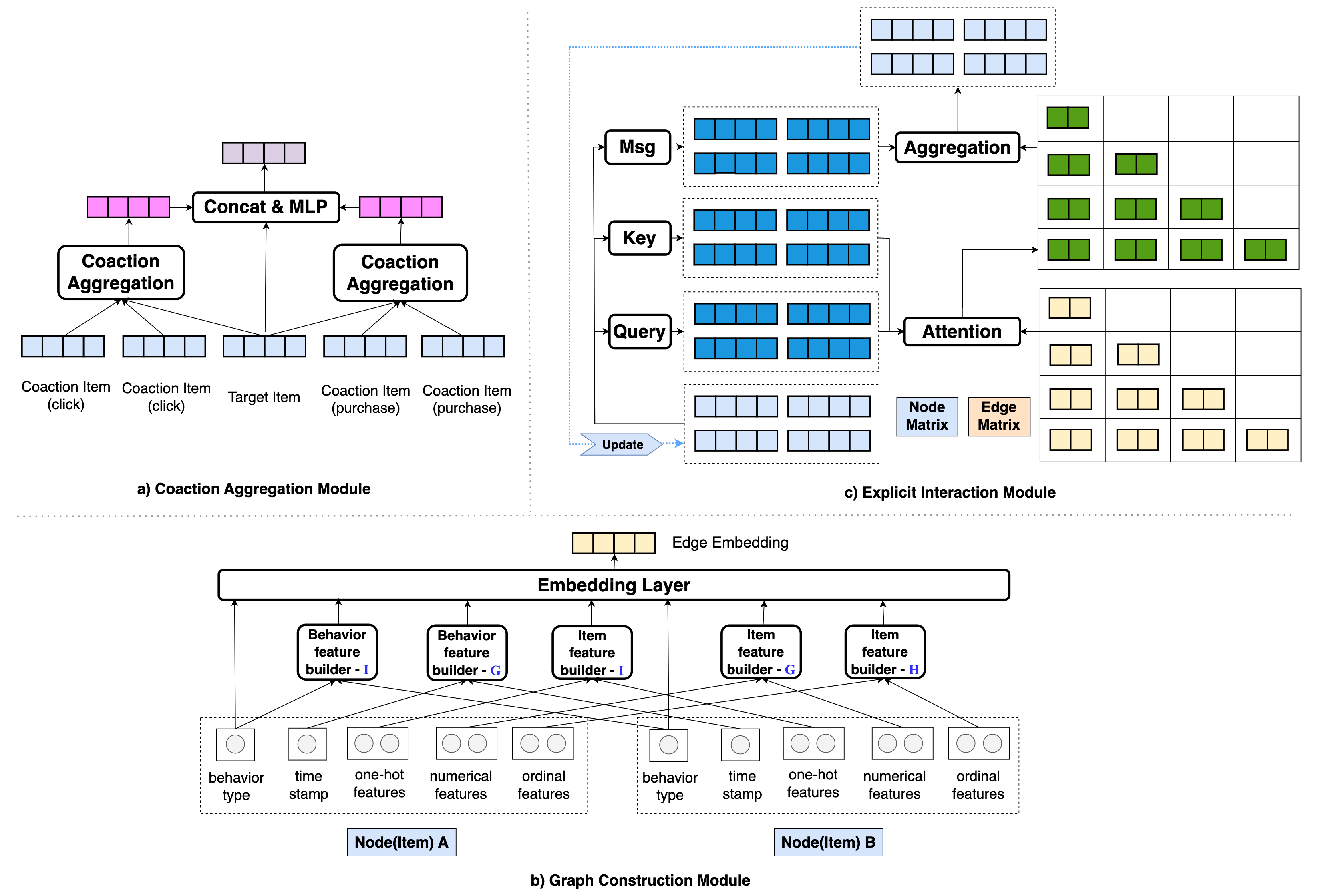}
    \end{minipage}
    \hfill
    \caption{Module details. Figure (a) shows the steps of Co-action Aggregation Module, (b) depict the process of edge embedding generation, (c) shows the steps of Explicit Interaction Module }
    \label{modules}
\end{figure*}

Figure~\ref{model} represents the overall architecture of our CAGR model in this paper, which is a typical two tower model. After a shared \textbf{Embedding Module}, each item $x$ that feed into the model will be converted into embedding representation $\textbf{e}$. Then we have a \textbf{Co-action Aggregation Module} on the Item Tower that aggregates the target item and its co-action items, generating a final item representation. Furthermore, we have a \textbf{Graph Construction Module} that converts the user's action sequence into a fully connected graph, which followed by an \textbf{Explicit Interaction Module} that learns the interaction relationships among user actions, finally we extract multiple interest representations of user actions through the \textbf{Interest Extraction Module}, generating multiple user embeddings. Figure~\ref{modules} shows the detailed steps for each specific module.

\subsection{Item Tower: Co-action Aggregation Module}

We can consider the full set of item-to-item relationships at eBay as a co-action graph. On the Item Tower of our CAGR model, we construct a localized subgraph by considering the co-action items of the neighbors of a target item. Figure ~\ref{graph} a) shows the relationship between the global item graph and the item subgraph. We then use a GNN to aggregate information within this subgraph and obtain an embedding representation for the target item. Inspired by~\cite{Hamilton17,Brody0Y22,VelickovicCCRLB18}, this module is basically composed of two steps. In step 1 we employ an attention mechanism to aggregate the neighboring nodes. Given each co-click item embedding $\textbf{e}_{qc_n}$, we calculate its attention $\boldsymbol{\alpha}_{qc_n}$ with target item $\textbf{e}_q$ using the method in~\cite{Brody0Y22}, and execute a weighted sum $\sum_{c_n}\boldsymbol{\alpha}_{qc_n}\textbf{e}_{qc_n}$ to get the aggregated co-click embedding, denoted as $\textbf{z}_{qc}$. For co-purchase items we can generate the corresponding embedding $\textbf{z}_{qp}$ as well. And in step 2 we concatenate the aggregated neighbors and transformed with a linear layer to obtain the final item embedding $\textbf{z}_q$, i.e. $ \textbf{z}_q = \textbf{W}_{zq} \cdot \textup{CONCAT}(\textbf{e}_q, \textbf{z}_{qc}, \textbf{z}_{qp}) + \textbf{b}_{zq}$ , where $\textbf{W}_{zq}$ and $\textbf{b}_{zq}$ are model weights.

\subsection{User Tower: Graph Construction Module}

In this module, we transform the user's sequence of behaviors into a fully connected graph. The graph is a directed graph where each node represents a user action $\{$click, add to watchlist, add to cart, purchase$\}$ on a specific eBay item. Each edge connects two user actions on eBay items to capture the temporal relationships between these actions based on the user's sequential behavior. We make the graph directed because early-occurring actions can affect later-occurring actions but not vice versa, so the direction of information aggregation should be one-way. Figure ~\ref{graph} b) gives an example of how to convert the sequence data into a sequence graph.

The main task in this module is to generate the representation for each node and edge on the graph. For each node we directly use its item embedding representation as node representation. For each edge we use $\textbf{E}(s_{u_i}, s_{u_j})$ to denote an edge which connects the $i$-th and $j$-th action of user $u$. To construct the edge information, we mainly build from pairwise relationship information. This type of information is calculated based on the features of a given node pair. Calculating pairwise level information provides an explicit way to represent the behavior interaction, which corresponds to our motivation. Here, we design several different types of functions based on the different characteristics of item features. For item $x$, we use $x^{(f)}$ to denote one of its features.
\begin{itemize}
    \item For one-hot features contained in items, such as item ID, category ID, parent category ID, seller ID, etc., we define the function $\textbf{I}$ to compare if they are the same.
    \begin{equation}
     \textbf{I}(x_a^{(f)}, x_b^{(f)}) = \begin{cases}
                                           1 & \text{if } x_a^{(f)} = x_b^{(f)} \\
                                           0 & \text{otherwise}
                                         \end{cases} 
    \end{equation}
    \item For numerical value features contained in items, such as processed price information, we define the function $\textbf{G}$ to measure the gap between them.
        \begin{equation}
    \textbf{G}(x_a^{(f)}, x_b^{(f)}) = x_b^{(f)} - x_a^{(f)}
    \end{equation}
    \item For ordinal features that can represent relative relationships, such as the seller's level, or bucketized numerical feature like price range level, we define the function $\textbf{H}$ to compare and return their order relationship.
    \begin{equation}
    \textbf{H}(x_a^{(f)}, x_b^{(f)}) = \text{sign}(x_b^{(f)} - x_a^{(f)})
    \end{equation}
         
\end{itemize}
Besides the pairwise information from item features, the behavior type $b_{u_i}$ and timestamp $t_{u_i}$ are considered as one-hot and numerical features separately, applying the above functions to them. We also convert $b_{u_i}$ and $b_{u_j}$ into embedding representations $\textbf{b}_{u_i}$ and $\textbf{b}_{u_j}$, and include them directly as part of the edge information.

Based on the above feature builder, for each item $x$ with $N$ features, we arrange the feature index and let features with the same type appeared in a row. $x^{(1)}, \ldots, x^{(g)}$ are one-hot features, $x^{(g+1)}, \ldots, x^{(h)}$ are numerical features, and $x^{(h+1)}, \ldots, x^{(N)}$ are ordinal features. Then we calculate the corresponding pairwise features based on above functions and concatenate all the results. Then, we can define the edge representation as:
\begin{equation}
    \begin{split}
     {\textbf{E}}(s_{u_i}, s_{u_j}) = & \text{CONCAT}(\textbf{b}_{u_i}, \textbf{b}_{u_j}, \textbf{I}(b_{u_i}, b_{u_j}), \textbf{G}(t_{u_i}, t_{u_j}), 
     \\
    & \textbf{I}(x_{u_i}^{(1)}, x_{u_j}^{(1)}), \ldots, \textbf{I}(x_{u_i}^{(g)}, x_{u_j}^{(g)}),  \\
    & \textbf{G}(x_{u_i}^{(g+1)}, x_{u_j}^{(g+1)}), \ldots, \textbf{G}(x_{u_i}^{(h)}, x_{u_j}^{(h)}), 
    \\
    & \textbf{H}(x_{u_i}^{(h+1)}, x_{u_j}^{(h+1)}), \ldots, \textbf{H}(x_{u_i}^{(N)}, x_{u_j}^{(N)}))
    \end{split}
\end{equation}

Since our graph is directed, the edge embedding representation only makes sense when the input $s_{u_i}$ and $s_{u_j}$ has the order $i \geq j$. For the case when $i < j$, we can simply define that the edge embedding is $\textbf{0}$. 

Finally we have constructed a sequence graph $\mathcal{G}_u = (\mathcal{V}_u, \mathcal{E}_u)$, where the set of nodes $\mathcal{V}_u$ and the set of edges $\mathcal{E}_u$ have the corresponding embedding representations as follows:
\begin{equation}
    \text{Emb}(\mathcal{V}_u) = (\textbf{e}_{u_1}, \textbf{e}_{u_2}, \ldots, \textbf{e}_{u_T})
\end{equation}
\begin{gather}    
    \text{Emb}(\mathcal{E}_u) = 
        \begin{bmatrix}
            \textbf{E}(s_{u_1}, s_{u_1}) & \textbf{E}(s_{u_1}, s_{u_2}) & \dots & \textbf{E}(s_{u_1}, s_{u_T}) \\
            \textbf{E}(s_{u_2}, s_{u_1}) & \textbf{E}(s_{u_2}, s_{u_2}) & \dots & \textbf{E}(s_{u_2}, s_{u_T}) \\
            \vdots & \vdots & \ddots & \vdots \\
            \textbf{E}(s_{u_T}, s_{u_1}) & \textbf{E}(s_{u_T}, s_{u_2}) & \dots & \textbf{E}(s_{u_T}, s_{u_T})
        \end{bmatrix}
\end{gather}

We note that each user has their own user graph, separate from other user's graph. So if we have 1 million training examples (corresponding to 1 million users), then we will have 1 million separate user graphs during model training.

\subsection{User Tower: Explicit Interaction Module}

In this module, we learn the explicit interaction between user behaviors based on the constructed sequence graph. The entire process is described in Algorithm \ref{eialgo}. The algorithm can be seen as an $L$-layer Graph Neural Network. Our initial input is obtained from the sequence graph output by the previous module. For a particular layer $l$, we perform an attention based graph layer for each node. The attention here is a customized attention mechanism that needs to consider both node and edge information. We borrow the attention calculation method from both GAT~(additive attention) and Transformer~(dot product attention). To be concrete, we generate the query embedding $\textbf{q}_i^{(l)}$ and key embedding $\textbf{k}_j^{(l)}$ based on the input node embedding $\textbf{h}_{u_i}^{(l)}$/$\textbf{h}_{u_j}^{(l)}$, and then use query embedding, key embedding, edge embedding, and the dot product of query and key as input when calculating attention. The customized attention is overall in an additive attention style, and for the input we include the edge embedding for explicit pairwise information, as well as the node dot product to ensure that the model can learn the bit-wise relevance from nodes.

\subsection{User Tower: Interest Extraction Module}

After the aforementioned steps, we have obtained a behavior representation matrix
$\textbf{H}_u = (\textbf{h}_{u_1}, \textbf{h}_{u_2}, \ldots, \textbf{h}_{u_T})$. Next, we employ a interest extraction module to extract the user's multiple interest representation. We adopt the exact self attentive approach in~\cite{Cen2020}, which is also used as Multi-Interest Extraction Layer in~\cite{ijcai2021p197}. The output of this module, which is the user's interest representation matrix, can be denoted as $\textbf{O}_u = (\textbf{O}_{u_1}, \textbf{O}_{u_2}, \ldots, \textbf{O}_{u_K})$, where $\textbf{O}_{u_i}$ is the representation vector for the $i$-th interest of the user, and $K$ is a preset parameter representing the number of interests. 

\begin{algorithm}
    \SetAlgoLined
    \LinesNumbered
    \KwIn{Sequence graph $\mathcal{G}_u$, which contains the node embeddings $\text{Emb}(\mathcal{V}_u)$ and edge embeddings $\text{Emb}(\mathcal{E}_u)$.
    }
    \KwOut{Enhanced sequence graph node embeddings $(\textbf{h}_{u_1}, \textbf{h}_{u_2}, \ldots, \textbf{h}_{u_T})$ }

    \textbf{Initialize}: $(\textbf{h}_{u_1}^{(0)}, \textbf{h}_{u_2}^{(0)}, \ldots, \textbf{h}_{u_T}^{(0)})$ = $(\textbf{e}_{u_1}, \textbf{e}_{u_2}, \ldots, \textbf{e}_{u_T})$ \\
        \For{$l \leftarrow 1$ \KwTo $L$}{
            \For{$i \leftarrow 1$ \KwTo $T$}{
                 $\textbf{q}_i^{(l)} = \textbf{W}_Q\textbf{h}_{u_i}^{(l-1)}$ \\
                 $\textbf{k}_i^{(l)} = \textbf{W}_K\textbf{h}_{u_i}^{(l-1)}$ \\
                 $\textbf{v}_i^{(l)} = \textbf{W}_V\textbf{h}_{u_i}^{(l-1)}$ \\
            }

            \For{$i \leftarrow 1$ \KwTo $T$}{
                \For{$j \leftarrow 1$ \KwTo $T$}{
                    $\textbf{A}_{ij}^{(l)} = \textbf{w}_{a}^{(l)} \cdot \textup{LeakyRelu}(\textbf{W}_{a}^{(l)} \cdot \textup{CONCAT}(\textbf{q}_i^{(l)}, \textbf{k}_j^{(l)},  \textbf{q}_i^{(l)} \cdot \textbf{k}_j^{(l)},  \textbf{E}(s_{u_i}, s_{u_j}))) $ \\
                }
            }

            $\textbf{M} = \text{CASUAL\_MASK}(\textbf{A})$ \\
            
            \For{$i \leftarrow 1$ \KwTo $T$}{
                $\textbf{A}_{i}^{(l)} = \text{SOFTMAX}(\textbf{A}_{i}^{(l)} + \textbf{M}) $ \\
                $\textbf{h}_{u_i}^{(l)} = \sum_{j}{} \textbf{A}_{ij}^{(l)} \textbf{v}_j^{(l)} + \textbf{h}_{u_i}^{(l-1)} $ \\
            }

        }
        \Return{$(\textbf{h}_{u_1}^{(L)}, \textbf{h}_{u_2}^{(L)}, \ldots, \textbf{h}_{u_T}^{(L)})$}\\

    \caption{Explicit Interaction Algorithm}
    \label{eialgo}
\end{algorithm}

\begin{table*} 
    \begin{tabular}{@{}c|cccccc|cccccc@{}}
        \toprule
                  & \multicolumn{6}{c|}{ Taobao } & \multicolumn{6}{c}{ eBay } \\
                  & R@20  & N@20 & H@20   & R@50  & N@50 & H@50  & R@20  & N@20 & H@20  & R@50  & N@50 & H@50  \\
        \midrule
        YouTube DNN     & 5.96 & 18.36 & 35.55 & 7.82 & 21.63 & 45.76 & 1.52 & 1.96 & 3.2 & 2.22 & 3.31 & 5.12   \\
        MIND            & 6.76 & 24.86 & 45.31 & 8.45 & 26.17 & 50.02 & 1.61 & 2.06 & 3.31 & 2.38 & 3.34 & 5.19   \\
        ComiRec         & 7.19 & 29.26 & 48.20 & 8.84 & 32.04 & 55.43 & 1.81 & 2.33 & 3.55 & 2.64 & 3.68 & 5.51  \\
        PIMIRec         & 7.57 & 31.03 & 50.51 & 9.36 & 34.21 & 57.88 & 1.89 & 2.52 & 3.84 & 2.79 & 3.92 & 6.01  \\
         \midrule
        CAGR         & \textbf{7.83} & \textbf{32.11} & \textbf{52.07} & \textbf{9.44} & \textbf{34.75} & \textbf{58.55} & \textbf{2.08} & \textbf{2.82} & \textbf{4.18} & \textbf{2.99} & \textbf{4.18} & \textbf{6.25} \\
        \bottomrule
    \end{tabular}
    \caption{Model performance on Taobao and eBay datasets (\%) demonstrating improved performance of the CAGR model.}
    \label{table:overall_perc}
\end{table*}

\begin{table}
    \begin{tabular}{@{}cccc@{}}
        \toprule
        \textbf{Dataset}  & \textbf{\#Users}  & \textbf{\#Items} & \textbf{\#Interactions} \\
        \midrule
        Taobao         & 0.97           & 1.71            & 95.27    \\
        eBay           & 46             & 278             & 1402     \\
        \bottomrule
    \end{tabular}
    \caption{Statistics of datasets (in millions)}
    \label{table:data_stat}
\end{table}

\subsection{Training Phase}

Now we have the interest representation matrix $\textbf{O}_u$ for a given user, as well as the representation vector $\textbf{z}_q$ for any target item. We follow the common approach in~\cite{Cen2020} and~\cite{ijcai2021p197} to pick the interest vector $\textbf{o}_{uq}^{(z)}$ that generate the highest dot product score which is the affinity score on the model architecture figure among the user interest vectors, i.e. $\textbf{o}_{uq}^{(z)} = \textbf{O}_u[\text{argmax}(\textbf{O}_u^\text{T}\textbf{z}_q)]$, and define the loss $\mathcal{L}_z$ as the sampled softmax loss where the target item is positive and random selected items are negatives. In addition, we compute the the score between the user vector and the embedding $\textbf{e}_q$ of the item itself as an auxiliary score, and select the activated interest vector $\textbf{o}_{uq}^{(e)} = \textbf{O}_u[\text{argmax}(\textbf{O}_u^\text{T}\textbf{e}_q)]$ in a similar manner. With $\textbf{o}_{uq}^{(e)}$ we define a corresponding loss $\mathcal{L}_e$ which is also the sampled softmax loss. While co-action items making full use of collaborative signals, they inevitably introduces some noise. Therefore, we define this auxiliary loss to allow us to intervene in the importance ratio between collaborative signals and the item itself during training, guiding the model to learn better embeddings. Thus, the final loss function is:
\begin{equation}
\label{eqn:loss}
    \mathcal{L} = \mathcal{L}_z + \lambda \mathcal{L}_e
\end{equation}
Where $\lambda$ is a hyperparameter to balance the ratio between $\mathcal{L}_z$ and $\mathcal{L}_e$.

The learned user embeddings $\textbf{O}_{u}$ and item embedding $\textbf{z}_q$ based on the given user $u$ and item $q$ can be used for calculate the user's score for the item:
\begin{equation}
    \hat{y}_{uq} = \max_{1 \leq i \leq K}(\textbf{O}_{u_i}^\text{T}\textbf{z}_q)
\end{equation}

\section{Offline Evaluation}\label{sec:experiments}

\noindent \textbf{Dataset and evaluation protocols.} We selected two datasets for evaluating recommendation performance, one public dataset and one industrial dataset. The statistics of the two datasets are shown in Table~\ref{table:data_stat}.

\begin{itemize}
    \item The Taobao user behavior dataset\footnote{https://tianchi.aliyun.com/dataset/649} is a publicly available dataset. To maintain consistency with related work, we refer to~\cite{Cen2020} and retain only items and users with at least 5 clicks, truncate the length of user behavior sequences to a maximum of 50, and use a user based split for training/validation/testing set. All four types of behaviors in the dataset are considered, but we limit the prediction target to click behavior.
    
    \item For the industrial dataset, we collects 8 days of eBay user behavior data, including click, add to watchlist, add to cart, and purchase. For this dataset, user behavior sequences are truncated at the length of 200. We perform a time-based split for training and testing, where the first 7 days of data are used for training, and the 8th day's data is used for testing. This split differs from the Taobao dataset as a time-based split ensures that the use of co-action behavior data as features does not encounter feature leakage during evaluation, and it aligns with the practical model training process in the industry.
\end{itemize}

\noindent \textbf{Baseline Models} Baseline models which we compare with our approach are YoutubeDNN~\cite{Covington2016}, MIND~\cite{Li2019}, ComiRec~\cite{Cen2020}, PIMIRec~\cite{ijcai2021p197}. Considering our CAGR method use multi-behavior signals, we also use the multi-behavior sequence when implementing baseline methods. Additionally, when conducting experiments on the Taobao dataset, our proposed method only utilizes the User Tower improvement, as the user based split in this dataset does not allow us to add item co-action data which will have the leakage problem between users. The number of interest is set to 4.


\noindent \textbf{Overall Performance.} Table~\ref{table:overall_perc} shows the results of all methods on Taobao and industrial datasets. We adopt the commonly used matching metrics, Recall@K, NDCG@K, and Hit Rate@K, with K = 20 and K = 50 for the metrics@K~\cite{Cen2020}. Our proposed CAGR method consistently outperforms the other baseline models on two datasets, which demonstrates that the proposed CAGR model can generate better user interest representations and improve recommendation performance.

\begin{table}
  \begin{tabular}{@{}c|ccc|ccc@{}}
        \toprule
                 & R@20  & N@20 & H@20  & R@50  & N@50 & H@50  \\
        \midrule
        $\text{CAGR}_{-c}$    & 2.01 & 2.69 & 4.07 & 2.89 & 4.10 & 6.16  \\
        $\text{CAGR}_{-p}$   & 2.07 & 2.80 & 4.14 & 2.97 & 4.16 & 6.23  \\
        $\text{CAGR}_{-a}$   & 1.98 & 2.66 & 4.03 & 2.86 & 4.07 & 6.12 \\
        \midrule
        $\text{CAGR}_{-e}$ & 1.98 & 2.68 & 4.03 & 2.89 & 4.08 & 6.16 \\
        $\text{CAGR}_{-g}$  & 1.90 & 2.52 & 3.86 & 2.81 & 3.94 & 6.05 \\
        \midrule
        CAGR  & \textbf{2.08} & \textbf{2.82} & \textbf{4.18} & \textbf{2.99} & \textbf{4.18} & \textbf{6.25} \\
        \bottomrule
    \end{tabular}
    \caption{Ablation study on eBay dataset (\%)}
    \label{table:ablation}
\end{table}

\begin{table}
 \begin{tabular}{@{}c|ccc|ccc@{}}
        \toprule
        & R@20  & N@20 & H@20  & R@50  & N@50 & H@50  \\
        \midrule
        ${\lambda}$ = 0.0 & 2.05 & 2.81 & 4.16 & 2.90 & 4.08 & 6.13  \\
        ${\lambda}$ = 0.1 & 2.07 & \textbf{2.82} & 4.17 & 2.91 & 4.13 & 6.18  \\
        ${\lambda}$ = 0.2 & \textbf{2.08} & \textbf{2.82} & \textbf{4.18} & \textbf{2.99} & \textbf{4.18} & \textbf{6.25}  \\
        ${\lambda}$ = 0.3 & 2.07 & \textbf{2.82} & \textbf{4.18} & 2.96 & 4.15 & 6.24  \\
        ${\lambda}$ = 0.4 & 2.05 & 2.81 & 4.14 & 2.92 & 4.14 & 6.22  \\
        \bottomrule
    \end{tabular}
    \caption{Impact of hyperparameter ${\lambda}$ in loss function on eBay dataset (\%)}
    \label{table:loss}

\end{table}

\begin{figure*}[t]
    \small
    \centering
    \begin{minipage}{1.00\textwidth}
        \includegraphics[width=\columnwidth]{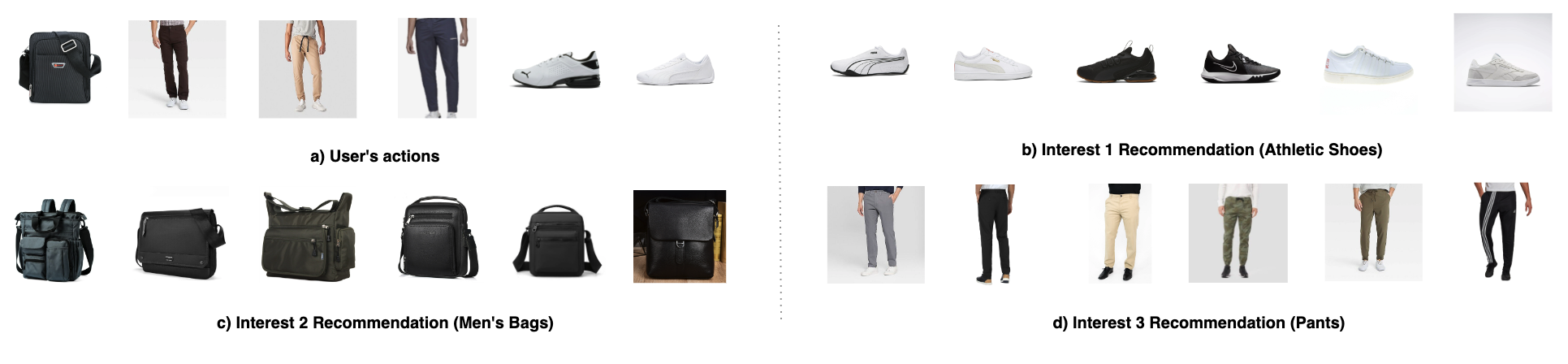}
    \end{minipage}
    \hfill
    \caption{Example of recommendations for an individual user, (a) shows the user’s historical interacted items, (b)(c)(d) shows the three interests we have captured for this user, with (b) representing the first interest on Athletic Shoes, (c) representing the second interest on Men's Bags, (d) representing the third interest on Pants.}
    \label{rec}
\end{figure*}

\noindent \textbf{Ablation Study of Co-action Aggregation.} We further investigated the effectiveness of Co-action Aggregation Module by separately removing co-click items($\text{CAGR}_{-c}$), co-purchase items($\text{CAGR}_{-p}$) and both of them($\text{CAGR}_{-a}$) on the industrial dataset. From the results in table \ref{table:ablation}, we can see that both click and purchase contribute to the model performance, with purchase contribute less than click probably because of the natural sparsity of this behavior.

\noindent \textbf{Ablation Study of Behavior Sequence Interaction.} We conduct some ablation experiments on the industrial dataset for the User Tower modules as well, including only remove the edge features ($\text{CAGR}_{-e}$) and remove the whole Graph Construct Module and Explicit Interaction Module ($\text{CAGR}_{-g}$), from table \ref{table:ablation} we can observe that removing the edge features leads to a decrease in the metrics, while the decrease in metrics becomes more significant if the whole graph is removed. This indicates that learning the sequence graph solely by aggregating node information contributes the model performance, which learns the interaction implicitly, but the explicit edge feature will bring a better performance.

\noindent \textbf{Impact of the Loss Function hyper-parameter $\lambda$.} Since we introduced two loss components from Eq~\ref{eqn:loss}, $\mathcal{L}_z$ and $\mathcal{L}_e$, we need to find an appropriate $\lambda$ to balance the proportion between them. A larger $\lambda$ indicates a higher proportion of $\mathcal{L}_e$, meaning that the importance of the target item itself is magnified while the importance of co-action data is weakened. When $\lambda$ decreases to 0, it means that we only consider the item embedding result with the fusion of co-action items. From the result Table~\ref{table:loss}, it can be observed that the model performs best when $\lambda$ is 0.2. This indicates that introducing the auxiliary loss $\mathcal{L}_e$ during training helps the model better balance the importance of the item itself and co-action data, thereby improving the model's performance. But overall $\mathcal{L}_e$ accounts for only 20\% of the weight, which indirectly confirms the importance of co-action data.



\section{Live Experiment and Online Serving}\label{sec:online-experiment}

In addition to offline evaluations, we also conducted a 14-day online A/B experiment. The control variant represents the production version deployed online, while the treatment variant incorporated our proposed CAGR method as an additional candidate generation recall set. We selected key performance metrics, including clicks, purchases, and revenue, as evaluation criteria. The results of the A/B experiment showed that our method led to a 1.82\% increase in clicks, a 2.16\% increase in purchases, and a 3.59\% increase in revenue over the production baseline.

In terms of the online serving of the model, since multi-interest models are better at mining users' long-term historical behaviors, under multi-recall online recommendation architecture which is a common industrial setting, we do not require real-time recommendation for this model. Instead we perform daily updates offline. For items, every day we collect attributes of all eligible recommended items and calculate the co-action data between these items. We then perform batch inference for all items to obtain the embedding set $\mathcal{Z}$ (this will constitute the KNN Item Index which will be used for retrieval). For users, we also have a daily batch inference based on the collected user historical behavior data, and generate the embedding set $\mathcal{O}$ for all users. Then, we generate each user's recommendation results based on the user's embedding set $\mathcal{O}$ and the item embedding set $\mathcal{Z}$. After that, for each user embedding $\textbf{O}_{u_i}$, a KNN search based on HNSW~\cite{Malkov2020} is performed in order to retrieve N items from the KNN Item Index, so we will get $K \cdot N$ items in total. A detailed engineering architecture diagram can be found in this blog~\cite{alacarte2022blog}, specifically refer to Figure 2. We then perform a ranking based on the retrieval score and select the final top-N result. Figure \ref{rec} shows an example of our recommendation results.

\section{Conclusions}\label{sec:conclusions}

In this work we proposed a multi-interest model that leverages a graph of co-action items of the target item, and enhanced User Tower with GNN based modules that learns pairwise interaction from the user behavior sequence. During the training phase, we introduced an auxiliary loss term to balance the model's emphasis on target item and its co-action items, enabling the model to better learn the representation. Both offline experiment and live A/B experiment shows the effectiveness of our proposed CAGR model over the baseline methods. In the future, we will consider adding more fine-grained co-action behaviors, such as incorporating user information into co-action events, and also expanding the training goals of our CAGR model beyond just click behavior to include other actions like add to cart, purchasing, and so on. At the same time, we will also consider integrating temporal graph-based techniques such as ~\cite{Zhou2021} and ~\cite{Fan2021}.

\begin{acks}
     We wanted to thank the generous support of Sathish Veeraraghavan, Bing Zhou, Jason Gong, Michelle Hwang, Sriganesh Madhvanath, Shawn Zhou, Yingji Pan, Menghan Wang, Liying Zheng, Jeff Kahn, Shuheng Li, Lili Weng, Ali Ismael, Dan Schonfeld for the help with the support of this project as well providing peer feedback.
\end{acks}

\bibliographystyle{ACM-Reference-Format}
\bibliography{references}

\end{document}